\documentclass{sig-alternate}
\usepackage{url}
\usepackage{fmtcount}

\begin{document}
%

\title{TwitterPaul: \\Extracting and Aggregating Twitter Predictions}

%
%
%
%
%

\numberofauthors{3} 
%
%
%


\author{
Naushad UzZaman\\
\affaddr{University of Rochester}\\
\affaddr{Rochester, NY, USA}\\
\email{naushad@cs.rochester.edu}
\alignauthor
Roi Blanco\\
       \affaddr{Yahoo! Research}\\
       \affaddr{Barcelona, Spain}\\
       \email{roi@yahoo-inc.com}
\alignauthor
Michael Matthews\\
\affaddr{Yahoo! Research}\\
\affaddr{Barcelona, Spain}\\
\email{mikemat@yahoo-inc.com}
}


\maketitle
\begin{abstract}
This paper introduces TwitterPaul, a system designed to make use of Social Media data to help to predict game outcomes for the 2010 FIFA World Cup tournament. 
To this end, we extracted over 538K mentions to football games from a large sample of tweets that occurred during the World Cup, and we classified into different types with a precision of up to 88\%. The different mentions were aggregated in order to make predictions about the outcomes of the actual games. We attempt to learn which Twitter users are accurate predictors and explore several techniques in order to exploit this information to make more accurate predictions. We compare our results to strong baselines and against the betting line (prediction market) and 
found that the quality of extractions is more important than the quantity, suggesting that high precision methods working on a medium-sized dataset are preferable over low precision methods that use a larger amount of data. Finally, by aggregating some classes of predictions, the system performance is close to the one of the betting line. 
Furthermore, we believe that this domain independent framework can help to predict other sports, elections, product release dates and 
other future events that people talk about in social media. 

\end{abstract}

\category{H.3.3}{Information Storage and Retrieval}{Information Search and Retrieval}
\terms{Algorithms, Experimentation}

\keywords{Twitter, retrieval, predictions}

\section{Introduction}

People are naturally interested in knowing future events. In our day to day life we try to forecast weather, natural disasters, the stock market, election results, sports outcome or anything that it is possible to bet on. 

Forecasting is a science where patterns are learned from historical data and used to predict the future: given a current situation, an individual issues a statement about what is going to happen next. Current methods are limited by the need for domain experts to understand the important parameters that govern a problem and the need for large amounts of historical data \cite{RePEc:eee:ecofch:1-01}. Forecasting has implications in financial decisions, risk management and investment decisions \cite{RePEc:tor:tecipa:tecipa-293}. 

Widespread use of social media started to change the scenario slightly. Some earlier work suggest that using social media might help to predict box office revenues \cite{AH10_BoxOffice}, polls from sentiment words extracted from Twitter  \cite{BO10_SentimentToPolls}, or even elections \cite{Tumasjan_Sprenger_Sandner_Welpe_2010} and the stock market \cite{JB10_BoxOffice}. The main rationale behind these approaches is that real people talking in social media have a direct impact on the outcome. For example, if someone tweets about \emph{Pirates of the Caribbean}, she might watch the movie, which indeed effects the box office revenue. The idea is very simple and seems to work in many domains. However, there are a number of caveats; for instance, it is important to consider the demographics problem, i.e. social media demographics might be biased and not reflect the actual demographics \cite{Dani11_Election,DBLP:journals/corr/abs-1204-6441} or the type of predictions we can make might be restricted by complexity of systems found in the social world \cite{DBLP:journals/corr/abs-1203-4642}. Furthermore, virality and network influence (word-of mouth) might affect significantly the information spread behavior of online social network users \cite{Leskovec:2006:DVM:1134707.1134732}.

On the other hand, the problem  of predicting the outcome of sports using social media is a challenge different than the ones just described: the chatter about sports in social media is unlikely to contribute to the outcome of the games. However, there might be some correlation between predictions made in social media and their outcomes in the real world. Many of those making predictions using social media will be basing their decisions on extensive analysis and expertise in the given sport. For instance, forecasting science might help to identify which team is stronger, if a key player is injured or out of form, the team's success rate in the recent past, whether if the team is playing in home ground or not, among many other parameters. Our goal is to develop techniques that harness the knowledge that is contained in social media: we do not need to analyze the game, but rather intelligently aggregate the analysis provided by the crowd. 
Understanding and analyzing the correlation between the crowd's voice and its ability to generalize to a domain in which it has no direct influence on the outcome is the main focus of this paper. People familiar with Football (Soccer) might be aware of \emph{Paul the Octopus}\footnote{\url{http://en.wikipedia.org/wiki/Paul_the_Octopus}} (26 January 2008 - 26 October 2010), who became a few-days celebrity wonder by making 8 correct predictions for World Cup Football 2010. The task in which we focus in this paper deals with predicting football games of the World Cup 2010 using Twitter data, and therefore we named the system after the infamous octopus as \emph{TwitterPaul}. Another key aspect of TwitterPaul is that the model is able to account for the past success rate of different users, and re-weight their predictions accordingly. 

TwitterPaul uses the extracted predictions in order to make guesses about the outcome of the game. We estimate the probability for the game outcome using three different methods: i) counting the predictions as votes, ii) counting the predictions as votes weighted with previous success rate, number of predictions, followers or friends, and iii) using machine learning algorithms to learn the weight that should be assigned in the model to a user's prediction. These results are compared with strong baselines and against the betting line (i.e., the prediction market), using data from an online betting site. Prediction markets are becoming increasingly popular; these are markets in which buyers and sellers can trade securities whose prices correspond to the predicted probability that a specific outcome will take place. In theory, no prediction method should be able to consistently outperform a proper prediction market, given that if someone could outperform the market they would have an incentive to make money in it \cite{watts:2011}. 


Among our findings we discovered that the quality of the extractions in large amounts is more important than the quantity, suggesting that high precision methods working on a medium-sized dataset are preferable to low precision methods that make use of  larger sizes of data. Furthermore, TwitterPaul is able to make predictions close to those of the prediction market, just using Social Media chatter. On a second round of experiments we simulate the performance of TwitterPaul in the betting market and examine whether its is possible to earn money using our system prediction or not. We found out that performance measured with root mean squared error (RMSE) is not correlated with the earnings from the betting market, and discuss why this is the case. 
The paper is organized as follows. Section \ref{sec:related} presents related work, Section \ref{sec:task} describes formally the task of predicting football games, Section \ref{sec:prediction} details the different Twitter-based prediction methods developed, which are evaluated in Section \ref{sec:evaluation}. The paper concludes in Section \ref{sec:conclusions}.
 \addtocounter{footnote}{1}

\begin{table*}[!ht]
\centering
\caption{Prediction Extraction Categories \label{table:categories}} 
\label{ExtractionExample}
\begin{tabular}{|c|l|}
\hline
Extraction & Example \\ 
\hline
Strong prediction & \#wc2010 \textit{ESP 2-0 NED}, \\
 &  \textit{i predict Portugal will win} for sure \\ 
\hline
Weak prediction  & Alemania \textit{2-0 Australia}$^{\decimal{footnote}}$, \textit{Spain wins} \\ 
\hline
Support & \textit{go spain}! you have to win it, \\
& \textit{i want brazil to win} \\
\hline 
Third person & psychic octopus \textit{paul predicts a german win} \\ 
\hline 
Prediction retweet & \textit{RT @foo ARG will win today} \\
\hline 
Question & @bar \textit{do you think brazil can win}\\
\hline 
Condition & \textit{if .. wins Bobbi Eden, Larissa Riquelme,} \\
& \textit{Maradona would entertain followers }\\ 
& \textit{if .. wins, i will cry/jump...} \\ 
& \textit{retweet and if USA wins win an iPhone} \\ 
\hline 
\end{tabular}
\end{table*}
\section{Related Work} 
\label{sec:related}

The huge growth in user generated content in recent years has led to a number of papers that employ social media information to make predictions about future events~\cite{DBLP:journals/corr/abs-1203-1647}. The contents of social media provide a mechanism to discover social structure and analyze action patterns qualitatively and quantitatively, and sometimes the ability to predict future human-related events.

Asur and Huberman \cite{AH10_BoxOffice} aim at forecasting the box office revenue by extracting tweets referring to movies, where the keywords present in the movie title serve as a query. They extracted 2.89M tweets from 1.2M users, which refer to 24 movies released over a period of three months, with the aim to predict the box office revenue generated by the movie in its opening weekend, using the tweets prior to their release. The prediction is implemented with a logistic regression model for which the training data is gathered using Amazon Mechanical Turk, and the system outperformed the predictions made by Hollywood Stock Exchange. They also experimented with a sentiment analysis tool with \textit{positive, negative and neutral} labels, and found that sentiments are not a strong signal compared to tweet-rate (without sentiment). On a similar line, Ming et al.~\cite{DBLP:journals/corr/abs-1203-4642} argue that Twitter users can be characteristically different from general users when compared to other online populations, and that this data cannot predict a movie box-office success. On the other hand, Mestyan et al.~\cite{DBLP:journals/corr/abs-1211.0970} built a predictive model for the financial success of movies based on the collective activity data of online users. They show that the popularity of a movie could be predicted well in advance by measuring and analyzing the activity level of editors and viewers of the corresponding entry to the movie in the Wikipedia.

A theme that has raised attention recently is the one of \emph{political predictions} \cite{Tumasjan_Sprenger_Sandner_Welpe_2010, 10.1109/HICSS.2012.607,mustafaraj_2010_political}. For instance,  O'Conner et al.~\cite{BO10_SentimentToPolls} aim at linking the sentiments found in Twitter to public opinions. They consider 1 billion tweets over the years 2008 and 2009 and gathered public opinion surveys from multiple polling organizations. They retrieved related messages by searching using just a few keywords, for instance \textit{obama} for presidential approval, \textit{obama and mccain} for election and \textit{economy, jobs and job} for consumer confidence. They calculated the day-to-day sentiment scores by counting positive and negative messages. Positive and negative words are defined by the subjectivity lexicon from OpinionFinder, a word list containing about 1,600 and 1,200 words marked as positive and negative, respectively; however, they do not use the lexicon distinctions between weak and strong words. A message is positive if it contains positive words, and negative if it contains negative words and a message could be both positive and negative. The sentiment score for a given day is calculated as the ratio of positive versus negative messages on that topic. Next they find the correlation between sentiment of Twitter messages and polling results that they gathered from multiple polling organizations. Their results varied across different datasets, but they found a correlation as high as 80\%. 

Tumasjan et al. \cite{Tumasjan_Sprenger_Sandner_Welpe_2010} investigate if online messages on Twitter validly mirror the off-line political sentiment. They prototyped the system for German federal election with around 104K tweets between 13th August to September 19, 2009 for the election taking place at September 27th, 2009. They collected all tweets that contained the names of either the 6 parties represented in the German parliament or selected prominent politicians of these parties who are regularly included in a weekly survey on the popularity of politicians. The authors employed LIWC2007 \cite{LIWC07}, a text analysis software developed to assess emotional, cognitive and structural components of text samples. This study focuses on 12 dimensions in order to profile political sentiment: future/past orientation, positive/negative emotions, sadness, anxiety, anger, tentativeness, certainty, work, achievement, and money. The system also automatically translates German tweets to English for sentiment extraction using LIWC2007. The results reported show that the share of attention the political parties receive on $\approx$100K tweets that were collected until one week before the elections reflects the elections result and  comes close to traditional election polls. They also claim the sentiment profiles of politicians and parties plausibly reflect many nuances of the election campaign. For example, the similar profiles of \emph{Angela Merkel} and \emph{Frank-Walter Steinmeier}, mirror the consensus-oriented political style of their grand coalition before this election.
In any case, the topic of whether social media can make accurate political prediction still remains disputed \cite{DBLP:journals/corr/abs-1204-6441}.

On a different stream of work, Filippova et al.~\cite{Filippova:2009:CES:1609067.1609094}, build company-specific summaries from a collection of financial news, in order to provide information on short-term stock trading. This work focuses on high-quality sentence retrieval rather than identifying and aggregating large quantities of low-quality predictions.

Bollen et al. \cite{JB10_BoxOffice} investigate whether measurements of collective mood states derived from large-scale Twitter feeds are correlated with the value of Dow Jones Industrial Average (DJIA) over time. They analyze the text content of daily Twitter feeds from February 28 to December 19th (9.8M tweets from 2.7M users) by two mood tracking tools, i. OpinionFinder that measures positive vs negative mood and ii. Google-Profile of Mood States (GPOMS) that measures mood in terms of 6 dimensions (Calm, Alert, Sure, Vital, Kind and Happy). In their experiment, they found that changes of public mood along with mood dimensions match shifts in DJIA values that occur 3-4 days later. They didn't find this effect for OpinionFinder's assessment of public mode in terms of positive vs negative, rather for the GPOMS dimension labeled \emph{Calm}. They trained a self-organized fuzzy neural network on the basis of past DJIA and public mood time series to predict DJIA closing values and their system has an accuracy of 87.6\% in predicting daily up and down changes in the closing values of DJIA. Despite the popularity of the study, it is worth noting that there are some inaccuracies in the model that would have biased the results, including data selection and testing of the model on the best data period available.\footnote{\url{http://sellthenews.tumblr.com/post/21067996377/noitdoesnot}\\\url{http://blog.someben.com/2011/05/sour-grapes-seven-reasons-why-that-twitter}\\\url{-prediction-model-is-cooked/}}

All of these works on Twitter perform simple text processing or sentiment analysis, usually by searching some relevant keywords or just matching the keyword in some dictionary. In our work for game prediction, we also use the Twitter data but for prediction extraction (sentiment analysis for our problem), we perform a sophisticated extraction method in order to understand which team the tweet is predicting as winner. 
\footnotetext[\value{footnote}]{\emph{Alemania} is the name for \emph{Germany} in Spanish, and assuming that the extractor is not able to match it to a team name, the prediction would be  \emph{2-0 Australia}, and classified as a weak prediction.}

\section{Task Description}
\label{sec:task}
The overall goal of our system is to investigate the feasibility of using Twitter to analyze and ultimately predict the outcome of future events. Formally, we will operate over a universe of games $g_i \in \mathcal{G}$, each one of them occurring at time $t_i$ and having a game outcome $o_i \in \mathcal{O} = \{team_1,draw,team_2\}$. We define the outcome $team_i$ as the event that the team $i$ wins the game, and $draw$ if no team wins the game. The system analyzes a set of tweets $T$, which it turns into \emph{game predictions} by a extraction process described in Section \ref{sec:extraction}. Predictions are tuples of the form $p = <u,g,o,c,t> \in \mathcal{P}$, in which $u$ refers to the user issuing the prediction, $c \in \{strong, weak, support, third,retweet,\\ question,  condition\}$ is a type of prediction (see Table \ref{table:categories}) and $t$ is the timestamp in which the prediction has been made.

TwitterPaul, focuses on using Twitter to predict game results for the World Cup Football/Soccer 2010, which took place in South Africa from June 11th to July 11th of 2010. 
In this case we extract $\mathcal{P}$ from a large sample of tweets that occurred in the 37 consecutive days starting one week before the start of the World Cup and ending just before the final game and $\mathcal{G}$ as the 64 games scheduled for the World Cup
where the event is defined for the two team playing and the time when the game took place.  
The system will estimate the individual probabilities of every outcome $o_i$ given a game $g_i$:
\begin{equation}
	\label{eq:prob}
	P(O=o_i|G=g_i,\mathcal{P}^i) \;,
\end{equation}
were $\mathcal{P}^i$ includes predictions \emph{prior} to the time of the game.

We have defined two different subtasks to address the problem
\begin{itemize}
\item \textbf{Extracting predictions}, or how to generate the set $\mathcal{P}$ out of a set of tweets. This requires identifying which one of the tweets refer to a prediction about an event in $\mathcal{G}$, and if so, extract the user stating the prediction $u_z$, the time in which the prediction has been made $t_i$, what is the predicted outcome $o_i$ and what is the category of the prediction $c_i$. In the case of Twitter, extracting $u_z$ and $t_i$ is straightforward given the meta-data found in the tweet, and the challenge is to identify $o_i$ and $c_i$ accurately. We describe this process in Section \ref{sec:extraction}.
\item \textbf{Predicting game outcomes}, or for a given event, predict the outcome of the event based on the set of extracted predictions. This corresponds to the estimation of the probability in Equation \ref{eq:prob}. We consider different alternatives, which we outline in Section \ref{sec:prediction} and evaluate in Section \ref{sec:evaluation}, against the real game outcome and against the prediction market.
\end{itemize}

Each of these subtasks is described in detail in the following sections.

\subsection{Extracting predictions from tweets using a Context Free Grammar}
\label{sec:extraction}
The primary focus of the prediction extraction task is to analyze the candidate tweets and extract from them all predictions made about the 64 candidate World Cup games.  
As a first step, we apply a simple first level filtering using team, soccer, prediction and support related keywords (only English tweets) to end up with total 16M+ (16,157,749) tweets for processing by the full extraction system.
After a pilot study in which a subset of the resulting tweets were analyzed, we determined that a better characterization of the data could be provided by classifying the predictions.  For example, we want to distinguish
between tweets that merely provide support for a team (\textit{I want spain to win}) from tweets that make a real game prediction (\textit{Spain will beat Holland}). In addition, since we
 profile users based on their past predictions, it is also important to distinguish between predictions made by the actual Twitter user and re-tweets or statements about somebody else's predictions. 
In all, we define seven categories that are summarized with examples in Table \ref{ExtractionExample}.

For each category, we extract the outcome of the game (either the winning team or draw) and optionally the score and crucially we map the predicted outcome directly to one of the 64 games on the World Cup schedule.

Unlike most of the approaches (\cite{AH10_BoxOffice, BO10_SentimentToPolls, Tumasjan_Sprenger_Sandner_Welpe_2010, JB10_BoxOffice}) that make use of social media data we approach the problem of prediction extraction using a Context Free Grammar (CFG) in order to capture the syntactic structure of the tweets
. The usage of the CFG provides a high-precision oriented technique for prediction extraction, given that all the information extracted will fall into a set of controlled structured syntactic patterns. This is feasible, given the character limitation of Twitter and the nature of prediction mentions, which are usually written in a limited number of ways - in this case applying the CFG rather helps to to capture the syntax instead of just words. 
A snippet of our CFG style grammar is shown below for a better understanding of the process.\footnote{Full grammar available upon request.} 
\begin{verbatim}
PREDICTION --> 
   OPINION_HOLDER PREDICT_WORD GAME_RESULT |
   GAME_RESULT |
   GAME_RESULT WORDS{0,1} FIRST_PERSON PREDICT_WORD

GAME_RESULT -->
   TEAM WORDS{0,2} MODAL_WORDS? WORDS{0,1} WIN_WORDS | 
   TEAM_SCORE | 
   TEAM AND TEAM WORDS{0,1} WIN_WORDS | 
   TEAM WORDS{0,1} WIN_WORDS (WORDS{0,1} TEAM)? | 
   WIN_WORDS CONNECTIVE_WORD TEAM 

TEAM_SCORE -->
   (GAME_SCORE WORDS{0,1} TEAM1 TEAM2)+ | 
   (TEAM1 VERSUS TEAM2 WORDS{0,1} GAME_SCORE)+ |              
   (GAME_SCORE WORDS{0,1} TEAM1 VERSUS TEAM2)+ | 
   (TEAM1 TEAM2 GAME_SCORE)+ | 
   (TEAM WORDS{0,1} WIN_WORDS? GAME_SCORE)+ | 

PREDICT_WORD --> (predict | bet | pick | ...)

MODAL_WORD --> (going to | will  |  ...)

WIN_WORD --> (win | do it  | make it  |  ...)

TEAM_WORD --> (arg | argentina | bra | brazil..) 
\end{verbatim}

These rules allow to extract the elements that are included in a prediction $p$. We start with a set of pre-defined words that refer to each one of the teams in $\mathcal{G}$, and a list of modal and prediction expressions. Every tweet in the data-set is parsed through the grammar and this process provides the user issuing the prediction $u_z$ (\texttt{OPINION\_HOLDER}), the predicted game outcome (which can be mapped effortlessly to $o_i$) and a set of modal words (\texttt{PREDICT\_WORDS}). The time of the prediction $t_i$ is taken from the tweet's meta-data. Note that one tweet might contain multiple predictions and that these grammar rules might match only a substring in the tweet.
The category of the prediction (see Table \ref{table:categories}) is determined using the presence of modal words in the tweet. We further consider a set of grammar rules using the words employed to make a prediction word to classify them into one of \emph{strong prediction}, \emph{weak prediction}, \emph{support tweets}, \emph{prediction from 3rd person}, \emph{retweet} and finally \emph{question or condition}. 

The more rules we have the better our coverage are in our domain. Same goes for terminal words like \texttt{PREDICT\_WORD}. In our grammar, \texttt{WORDS\{n,m\}} means at least $n$ words and at most $m$ words. This allows to handle different words in between. This simple addition of \texttt{WORDS\{n,m\}} made the grammar robust enough to handle unstructured sentences in the social media. We can also extract multiple predictions in the same format in a single tweet, e.g. \textit{Prediction for next matches: ARG 2 : 1 GER, ESP 1 : 0 NED}. However, if this tweet had one prediction in \texttt{team score : score team} format and another in \texttt{team score - score team} format, then we limit the extraction to only one of them. 
As an example on how the CFG captures sentences with grammar rules, consider, \textit{Spain is going to win}, \textit{Brazil will win} and \textit{Germany will make it}, which are all instances of 
\texttt{TEAM WORDS\{0,2\} MODAL\_WORDS? WORDS\{0,1\} WIN\_WORDS}. \\
Although our grammar were meant to extract English predictions only, the CFG will extract many predictions in other languages provided that the team name is in the set of \texttt{TEAM\_WORD}. This happens a non-negligible number of times, because many of the country mentions are written in 3-letter team codes, even though the rest of the tweet is in a different language. One example of extracting such tweet is, \textit{URU - NED ik zeg 1-3}. Here, \textit{ik zeg} is Dutch, which means ``i say''. 

Finally, in order to assign the prediction to one of the games in $g \in \mathcal{G}$ we require that the time extracted is lower than the time of $g$. In case that two teams play more than one game, we assign the one that took place closer to the prediction date. 


\section{Prediction of game outcomes}
\label{sec:prediction}
Predicting the outcome of the game is far from trivial. When people want to forecast how a game will end, they analyze which team is stronger, if the team is successful in the recent games, if the team is playing in the home ground, if a key player is injured, if a key player is out of form and many other parameters. The main idea of using social media to predict the game is to let people do all these analysis and write about it in the social media, so that we can aggregate their predictions to make a guess on the outcome of the game. Therefore, the goal is as follows: given predictions for a game and the predictors' previous history (context), predict the outcome of the game. 

\subsection{Prediction Market or Betting Line Baseline} As a reference, we used the betting line baseline which is an upper bound for all systems. Some theories in social sciences argue that it is not possible to beat the prediction market. The rationale is that the market represents a kind of the best knowledge we have before the game takes place, since people are putting their money for it, so they do research before predicting a winner \cite{watts:2011}. We retrieved historical betting records from the {Odds Portal\footnote{\url{http://www.oddsportal.com/soccer/africa/world-cup-2010/results/}}} website, which can be converted to probabilities.\footnote{\url{http://probetbonus.com/betting-theory/odds-probability}}

\subsection{Naive methods} 
\label{sec:baselines}
In this section we describe several methods to predict the probability of a team winning, as an aggregate over the whole extracted prediction set $\mathcal{P}$.

\textbf{Naive  Count}: This method is influenced from other social media based prediction methods (e.g. movie box revenue \cite{AH10_BoxOffice, Goel10}, or elections \cite{Tumasjan_Sprenger_Sandner_Welpe_2010}), where if there is a mention of an entity in a message, then it counts as vote for that entity. We run our team name extraction module and extract the team names from the tweets. Given $team_x$ and the tweet time for a tweet, if $team_x$ is playing in next $n$ matches ($n$ = 7), then we count one vote for $team_x$. In this case, if a tweet has multiple mention of a team, like a support tweet, \textit{Brazil! Brazil! Brazil! Brazil! Brazil!}, then we count only one vote for Brazil. 


\textbf{Coin Flip}: This baseline method gives 50\% probability for both teams, i.e. no probability for $draw$. 

\textbf{Team Ranking}: The FIFA compiles monthly a world ranking system for men's national teams, based on their recent game results, where the most successful teams are ranked higher. We have compiled this ranking prior to the World Cup in order to elaborate two different methods. The first one (team ranking)	will always predict a win for higher ranked team with probability $1$.\footnote{\url{http://www.fifa.com/worldranking/rankingtable/index.html}} 

\textbf{Team Ranking Odds}: We smooth the probability of a team winning ($team_1$) against another team ($team_2$) conditioned on both its opponent ranking and its own ranking, using the following formula: 
\begin{equation} 
1 - \frac{r(team_1)}{r(team_1)+r(team_2)}	\\
  		 =  \frac{r(team_2)}{r(team_1)+r(team_2)}	\;,
\end{equation} 
where $r(i)$ represents the rank of team $i$.

These last two baselines embody knowledge of experts in the field, assembled out of a sophisticated scoring mechanism, and as such, they are considered strong prediction methods.

\subsection{TwitterPaul}
In order to compute the probability distribution $P(\cdot)$ TwitterPaul differentiates between different types of users and different types of predictions by assigning different weights $w_i$ to each one of them.
We aggregate the different extracted predictions and compute a probability distribution for the outcomes of each game as:
\begin{align}
\label{ProbDistribution}
	P(O=o|G=g,\mathcal{P}) \approx \\
	\approx \frac{1}{|Z(g)|}\sum_{p_z = (u_z,g,o,c_z,t_z) \in \mathcal{P}} P(O=o|c_z,u_z) \\
	\approx \frac{1}{|Z(g)|}\sum_{p_z = (u_z,g,o,c,t_z) \in \mathcal{P}} P(O=o|c_z)P(O=o|u_z)\\
	= \frac{1}{|Z(g)|} \sum_{p_z = (u_z,g,o,c_z,t) \in \mathcal{P}} w_{c_z} \cdot w_{u_z}\;,
\end{align}
where $|Z(g)|$ counts the tuples that contain $g$ in $\mathcal{P}$, and where we will learn the user weights $w_{u_z}$ and prediction class weights $w_{c_z}$, which are normalized to lie between 0 and 1. These weights correspond to the probabilities $P(O=o|u_z)$ and $P(O=o|c_z)$ respectively. 
In the Evaluation section we will determine the contribution of the different prediction classes by setting $w_{c_z}=1$ for one class and $w_{c_z}=0$ for the rest of them, and further combining them by a simple mixture (\emph{all extractions}) by setting  ($\forall i,j$)  $w_{c_i}=w_{c_j}$. 

\subsection{Prediction Using Historical features} 
\label{sec:trust}



We now explore a number of user-dependent methods to assign the weights $w_{u_z}$, this is, given a user $u_z$ how likely she is to state a correct prediction. Note that we omit from this estimation structural information such as one user consistently stating the right predictions for the same team. Some of these features are simple and can be extracted automatically from the collection of predictions $\mathcal{P}$, like the total number of predictions made by a user before time $t$:

\begin{equation}
	\label{eq:aggregate}
		A(u_z,t) := \sum_{(u_z,g,o,c,t_p) \in \mathcal{P}} I\{t_p<t\}\;,
\end{equation}
where $I\{.\}$ is the indicator function.  Other features extracted from metadata include the number of followers of the user and number of friends of the user issuing the prediction. 
We describe some features that address the idea that users issuing the right predictions in the past are more trustworthy than other with random patterns of success. This idea is engineered using two different sources of information, namely, the \emph{previous success rate} and the \emph{number of predictions issued}.

For users $u_z$ with several predictions in $\mathcal{P}$, we can aggregate their success rate at time $t$ as:
\begin{equation}
	\label{eq:sum}
	S(u_z,t) := \frac{1}{|A(u_z,t)|} \sum_{(u_z,g,o_g,c,t_p) \in \mathcal{P}}  I\{ t_p < t, o_g = \hat{o}_g\} \; ,
\end{equation}
where $\hat{o}_g$ is the real outcome of game $g$. Equation \ref{eq:sum} stands for the number of correct predictions stated from $u_z$ before time $t$.


Next we capture the history of making predictions for a particular user before time $t$ as:


\newcommand{\argmax}{\operatornamewithlimits{argmax}}

\begin{equation}
	\label{eq:normalized}
	N(u_z,t) := \frac{|A(u_z,t)|}{\argmax_{k}\sum_{(u_k,g,o,c,t_p) \in \mathcal{P}} I\{t_p<t\}}\
\end{equation}
where $A(u_z,t)$ captures the total number of predictions made by a user before time $t$ and the denominator captures the maximum number of predictions made by any user before time $t$. 



The final trustworthiness score is a convex combination of Equations \ref{eq:sum} and \ref{eq:normalized}:
\begin{equation}
	\label{eq:trustworthy}
	T(u_z,t) = \lambda \cdot N(u_z,t) + (1-\lambda) \cdot S(u_z,t)\;,
\end{equation}
where $\lambda \in [0,1]$.
\subsubsection{Learned weights}
\label{sec:ml}

We also experimented with a machine learning approach to set the weights $w_{u_z}$, using logistic linear regression and the previous outcomes of the predictions for a given user. The goal is to assign a weight for a user issuing her $n^{th}$ prediction at time $t_n$. The loss function will be assembled over the previous $n-1$ predictions of the user and if they were correct or not.\footnote{The loss function quantifies the empirical error on the \emph{outcome} of the prediction} The features the model uses as input are the first ${n-1}$ prediction outcomes, the number of predictions made, and the number of friends and followers at time $t_n$.  
The output of the logistic regression model, which computes the conditional probability of the user issuing the right prediction for the game, is assigned to the user predictions weight. The experimental section also reports on learning the user weights with Support Vector Machine models. 

\begin{table*}
\centering
\caption{Extraction performance for different prediction classes. The proportions are estimated using a confidence interval with a level of $95\%$. The total number of predictions extracted in the data-set adds up to 538K.}
\label{ExtractionPerformance}

\begin{tabular}{|l|c|c|c|c|c|}
\hline
Extraction & Precision & Recall & F & Total \\
 &  &  & & instances\\
\hline
\hline
Strong Prediction & 88.6\% & 78.6\% & 83.3 & 117K- 170K\\
\hline
Weak Prediction & 40.8\% & 95.2\% & 57.1 & 148K - 206K\\
\hline
All Predictions & 69.3\% & 93.1\% & 79.4 & 286K - 347K\\
\hline
Support Tweet & 84.3\% & 73.7\% & 78.6   & 70K - 117K\\
\hline
Question/Condition & 84.6\% & 73.3\% & 78.5 & 31K - 67K\\
\hline
Third person & 61.9\% & 76.4\% & 68.4   & 24K - 57K\\
\hline
Prediction retweet & 75.0\% & 88.2\% & 81.1 & 23K - 55K\\
\hline
\end{tabular}

\end{table*}

\section{Evaluation}
\label{sec:evaluation}
\subsection{Evaluating prediction classification accuracy}

The system extracted 538K predictions pertaining to football games during the World Cup that occurred from June 11 to July 11, 2010. 
We ran our prediction extraction module on randomly selected 8776 tweets for evaluation, out of which the tools extracted 300 predictions belonging to different categories ($\approx$3.42\%). Those extractions were manually hand-labelled and assigned to one category, and used as ground truth to evaluate the performance of the CFG. Table \ref{ExtractionPerformance} presents the macro and micro performance over the different prediction categories portrayed in Table~\ref{ExtractionExample}.

The total number of predictions of each class are reported as confidence intervals at a 95\% confidence level, using the estimate of a population proportion without continuity correction. 

\begin{figure}[htb]
\centering
\includegraphics[scale=.30]{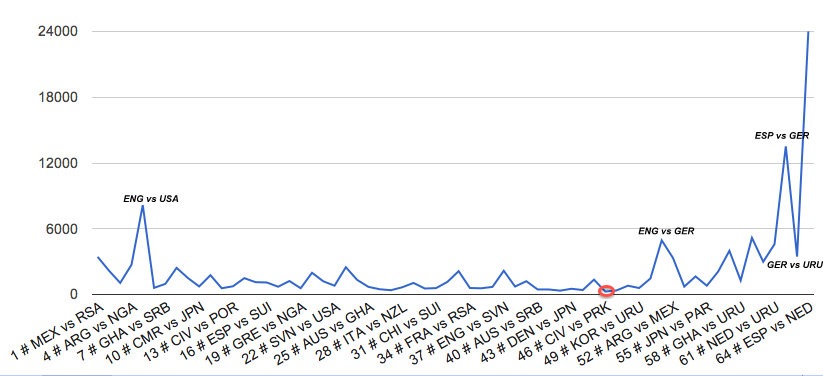}
\caption{Distribution of the \emph{strong predictions} extracted.}
\label{game_distribution}
\end{figure}

An extraction is considered correct if it is a game prediction, if the winning team is detected correctly and if it has been assigned to the right prediction class. The performance varies across classes. The overall precision and recall values are high, and results exhibit the typical precision-recall trade-off across different types of predictions, reflecting the variability of the CFG in detecting different classes.  Precision is lower for more ambiguous classes like \emph{weak prediction} or \emph{third person} for which the CFG fails to cover some of the classes or modal words. On the other hand the \emph{strong prediction} class exhibits higher performance figures.
The distribution of the extracted strong predictions is shown in Figure \ref{game_distribution}. The fact that the CFG focuses on English tweets might be the reason why there are high peaks for games of English speaking nations.

\subsection{Metrics}
\label{sec:rmse}
There are two intuitions that are important to be reflected in an evaluation metric when it comes to games prediction.
First, if one system predicts 51\% chance of win for a team and another system predicts 90\% chance of win for the same team and if that team wins, we expect the second system to be considered the 
better system.  Second, if one system predicts $team_1$ to win and a second system predicted a draw and $team_2$ actually wins, we again expect the second system to be considered superior. To capture these
intuitions, we use root mean square error (RMSE) as our evaluation metric, which Goel et al. \cite{Goel10} also used for evaluating the prediction tool that predicts game outcomes. 





RMSE quantifies the average difference between predicted and actual outcomes as
\begin{equation}
	 RMSE = \sqrt{\frac{1}{n}\sum_{i=1}^n \left( p_i - X_i\right)^2} \;,
\end{equation}
where $p_i$ is the predicted outcome for the game $g_i$, which we defined in equation (\ref{ProbDistribution}), and $X_i$ is the actual outcome.

In this case, $p_i$ will reflect the probability of winning for $team_{1}$, i.e. $P(O=team_{1}|G=g_i,\mathcal{P})$, and we define $X_i \in \{1,0.5,0\}$, which corresponds to the winner of the game being $team_1$, $draw$ or $team_2$ respectively.

Additionally, for the world cup football there are a few draw games ($\approx$20\% in World Cup 2010), since the later stage games are all knock-out games and the earlier games are mostly between uneven teams. Hence, we divide the $draw$ probabilities equally to both teams as follows: 

\begin{equation}
	 p_i =  P(O=team_{1}|G=g_i,\mathcal{P}) + \frac{1}{2} * P(O=draw|G=g_i,\mathcal{P})
\end{equation}

However, for league games it will be worthy to keep the $draw$ probabilities. 

We also evaluate our predictions against the betting-line predictions. In that case, $X_i$ is the probability of $team_1$'s winning, calculated from the odds for $team_1$'s winning.



\subsection{Baseline Performance}
\label{sec:baselineperformance}
Table \ref{BaselinePerformance} reports the performance for all the baselines described in Section~\ref{sec:baselines}. We observe in our data that the naive count baseline, which was influenced by other social media based systems by considering name mention as a vote, is better than the ordinal ranking and better than the coin flip baseline. However, if we modify the team ranking method by converting the ranking to a probability it outperforms all other baselines including the naive count. Note that we report both the performance using RMSE against the real outcome of the game and against the betting line, which reflect the closeness of the predictions made from the method with respect to those of the prediction market.

\begin{table}[htdp]
\caption{Baseline Performances}
\label{BaselinePerformance}
\begin{center}
\begin{tabular}{|l|c|c|}
\hline
Baseline & RMSE against & RMSE against\\
& actual result & prediction market \\
\hline
\hline
Ordinal team ranking & 0.5376 & 0.3589 \\
\hline
Coin flip & 0.4419 & 0.2416 \\
\hline
Naive count & 0.4322 & 0.1772 \\
\hline
\hline
\hline
Probability team &  &\\
 ranking & 0.4215 & 0.1614 \\
\hline
Betting line & 0.3959 & 0.0000 \\
\hline

\end{tabular}
\end{center}
\end{table}%

\subsection{TwitterPaul Performance}
\label{sec:twitterpaulperformance}

\textit{Performance of different types of predictions}. We firstly calculate the performance when we make use of different classes of predictions, which is reported in Table \ref{PerformanceDifferentExtraction}.

The rows in the table indicate for the methods that employ the different predictions types, i.e., \emph{strong predictions}, which have been extracted with very high precision ($>88\%$), \emph{only predictions} which contains both strong and weak predictions and finally \emph{all prediction}, which contains all the extractions, including predictions, support, retweet predictions, third person predictions and others. 
The data shows that when the method makes use of the strong predictions class this results in the lowest RMSE against both the actual game outcome and the prediction market. This happens in spite of the higher number of instances of other extraction classes. In this case, a larger quantity of data was less helpful since its inclusion ended up compromising quality. This effect is   more noticeable when comparing against the prediction market. 
As per this experiment, we conclude from our data that high quality extraction combined with good quantity (total of 150K) outperforms lower accurate predictions in larger quantities. However, it is matter of experimentation to find what is the balance between quantity and quality. 

Other classes of predictions resulted in a lower performance. The support tweets for instance performed poorer than the naive count baseline. This case is interesting, as the extraction precision of this class is 84.3\%; however, the amount of data available does not suffice to conclude how well the support tweets can be trusted for predicting game outcome. 

\begin{table*}[htdp]
\caption{Performance when the system uses different types of predictions. \emph{All extractions} stands for the performance of all the types of predictions combined. Performance is computed using RMSE against the actual outcome of the game and against the betting line.}
\label{PerformanceDifferentExtraction}
\begin{center}
\begin{tabular}{|l|c|c|c|c|}
\hline
Extraction & RMSE against & RMSE against & Total & Precision\\
& actual result & prediction market &  instances & \\
\hline
\hline
Strong prediction & 0.4197 & 0.1159 & 150K & 88.6\%\\
\hline
Only predictions & 0.4245 & 0.1409 & 323K & 69.3\%\\
\hline
All extractions & 0.4260 & 0.1553 & 538K & 30.5\%*\\
\hline
\hline
Team ranking & 0.4215 & 0.1614 & &\\
\hline
Betting & 0.3959 & 0.0000 & &\\
\hline
\hline
\hline
Support & 0.4416 & 0.2217 & 97K & 84.3\%\\
\hline 
Naive count & 0.4322 & 0.1772 & 10.9M  &\\
\hline
\end{tabular}
\end{center}
\end{table*}%

Having observed that the performance using the \emph{strong prediction} class is better than using other extraction types (Table \ref{PerformanceDifferentExtraction}), for rest of the experiments we will only consider 150K strong predictions. 

\textit{Performance of different weighting schemes}. In the next experiment, instead of considering each prediction as one full vote, we will weight the votes, i.e. we give higher weights to the vote from people with a more accurate prediction history, as described in Section~\ref{sec:trust}.

\begin{table}[htdp]
\caption{Performance with Weighted Count}
\label{PerformanceWeightedCount}
\begin{center}
\begin{tabular}{|l|c|c|}
\hline
Weighted  & RMSE against & RMSE against\\
Parameter & actual result & prediction market\\
\hline
\hline
1-RMSE & 0.4154 & 0.1115\\
\hline
previous predictions & 0.4131 & 0.1204\\
\hline
number of followers & 0.4183 & 0.1148 \\
\hline
number of friends & 0.4158 & 0.1124 \\
\hline
all four averaged & 0.4169 & 0.1104 \\
\hline 
$Trustworthiness$ & 0.4147 & 0.1131 \\
\hline
\hline
non-weighted & {0.4197} & {0.1159} \\
\hline
Betting & {0.3959} & 0.0000 \\
\hline
\end{tabular}
\end{center}
\end{table}%

To weight the predictions, we consider different features, such as the user's previous performance (calculated using the $1-RMSE$ score of her predictions), the number of previous predictions made by the users, her number of followers and friends, and so on. Those features are either used on their own as weights, plugged directly into Equation \ref{ProbDistribution}. 
Table \ref{PerformanceWeightedCount} reports the performance for the raw features, where the row names are self-explanatory. The trustworthy score was defined earlier in Section \ref{sec:trust}, which is the weighted average between $1-RMSE$ and $number$ $of$ $predictions$ (Equation~\ref{eq:trustworthy}), and $\lambda$ is learned on the available historical data for the game ($\lambda=0.5$ otherwise).

The outcome of the experiment reflects that in this domain, none of these parameters had any significant edge over the others or the unweighted solution, when comparing the performance against the betting line. On the other hand, some other features like number of friends and followers did not perform worse and at the same time features like previous history ($1-RMSE$) and number of previous predictions did not perform significantly better. When comparing the performance against the actual game result, the usage of historical data (\emph{previous predictions}, \emph{trust}) makes a larger difference when compared to the non-weighted method and the one that uses the team rankings. 

In order to gain a better understanding about the role of historical data we learned a user weight combining all the features using logistic regression and SVMs. We report the performance of these methods in Table \ref{PerformanceLogisticRegression}.

\begin{table}[htdp]
\caption{Performance with Learned Weights}
\label{PerformanceLogisticRegression}
\begin{center}
\begin{tabular}{|l|c|c|}
\hline
Regression Model & RMSE against & RMSE against\\
 & actual result & prediction market\\
\hline
\hline
Logistic Regression & 0.4183 & 0.1117\\
\hline
SMO & 0.4198 & 0.1242\\
\hline
\hline
\hline
trust & {0.4147} & 0.1131 \\
\hline
non-weighted & {0.4197} & {0.1159} \\
\hline
Betting & {0.3959} & 0.0000 \\
\hline
\end{tabular}
\end{center}
\end{table}%

Results compared to the actual game results are roughly similar to those of the unweighted solution, when all predictors are given equal weight, and worse than the \emph{trust} method. On the other hand, the logistic regression-learned weights produce the lowest RMSE against the prediction market, signifying that the method is able to leverage the signal in Twitter data to down- or up-weight some of the users' predictions. In every case, the machine learned weights still outperform all the baselines, including those based on previous team rankings. 

\begin{figure}[htb]
\centering
\includegraphics[scale=0.4]{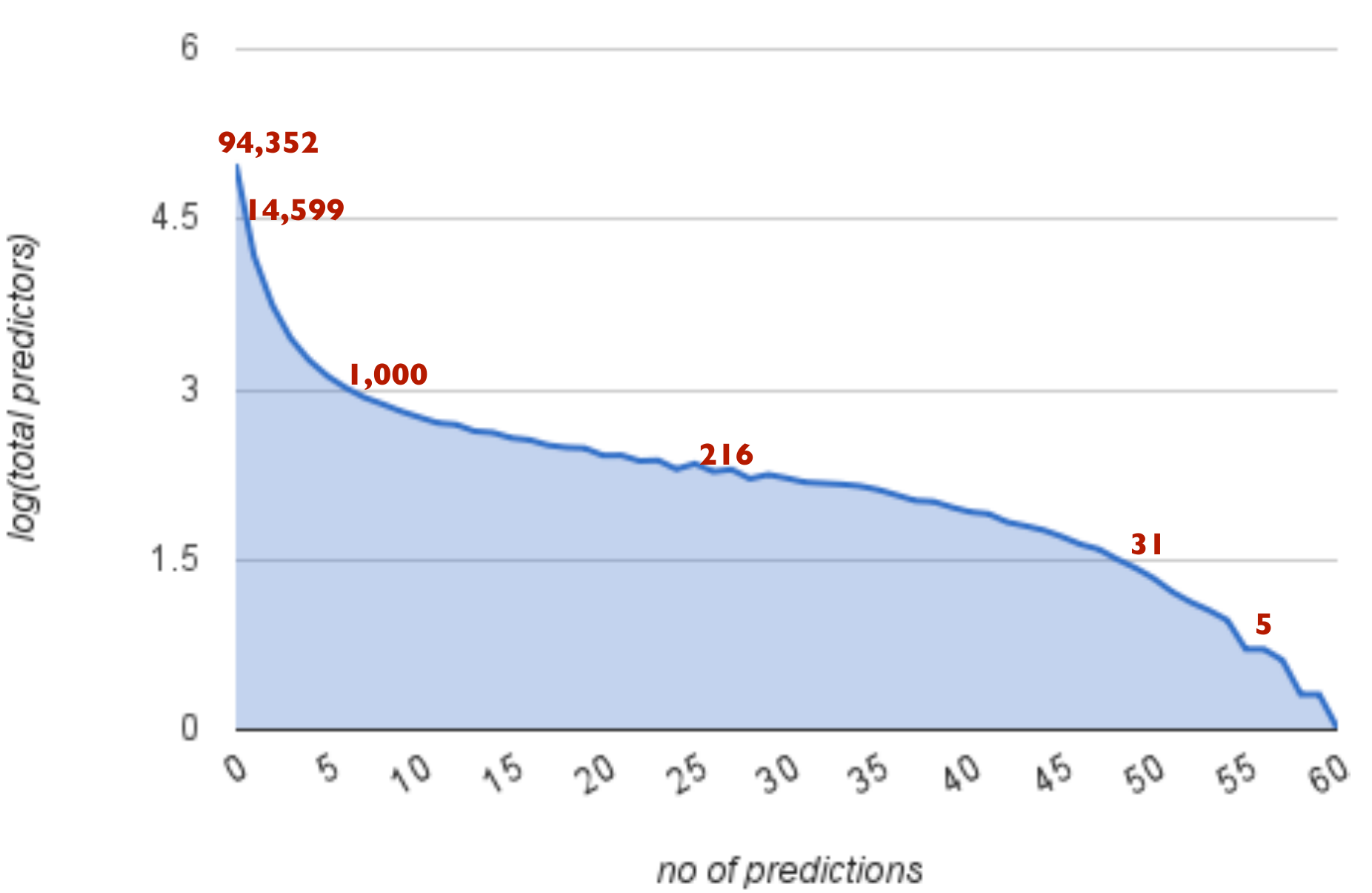}
\caption{Prediction Distribution}
\label{prediction_distribution}
\end{figure}

To understand better the effect of historical features in the model, we performed an error analysis. One observation is that most of the users lack a large history in the first place and the domain selected accounts for 64 games. Figure \ref{prediction_distribution} plots the number of previous predictions made by a user in the x-axis (could be at most 63) and the logarithm (base 10) of the total number of users making those predictions in y-axis in. As it can be observed, the curve exhibits a long tail.

Out of the total number of users, around 94K predictors make a single prediction, around 1K users made 5 (previous) predictions, and only 31 users made 50 (previous) predictions. This could account for one of the reasons why history in this domain failed to improve the performance. Further, given that weighting the performance also did not decrease the performance, those methods could be valuable for other tasks, for instance to discard spammers. 

In any case, it is known that experts with better history are usually unsuccessful at beating the performance of the average of a larger crowd \cite{Dani06_UAI}. Given the evidence available from historical data, this fact also holds in our domain, since we always have a group of new user predicting games much larger in number than few selected experts. 

\textit{Can we predict the game?}. Given the performance of the different methods, the question is, to what extent these methods can actually predict the games. The betting line or prediction market is the best knowledge available \cite{watts:2011} before the game, since people put real money and hence they do research before deciding which team to support for.
The outcome of the experiments show that TwitterPaul performs better than every other baseline (including methods very difficult to beat), but slightly worse than the betting line.
The finding that the strongest method shows insignificant differences with the prediction market goes in the same line as Goel et al.~\cite{Goel10}, who also found the betting line outperformed every other system, although the difference was insignificant.



\textit{Porting to other domains}. The prototype developed for the soccer domain is able to make predictions that correlate with the actual game outcome and which show an even higher correlation with the prediction market, even though people who are predicting are not participating in the actual game outcome. As mentioned earlier, current prediction tools require deep domain understanding to extract a handful of useful features to aggregate, and to engineer those to perform accurate predictions. Interestingly, in order to use social media chatter to make predictions one has to only create a domain-dependent extraction module and then in order to make predictions one could operate using the techniques described. Therefore, we believe in the ability of using this same framework for predicting the outcome of other sports, elections or other events that people talk about in social media.
\begin{table}[htdp]
\begin{center}
\caption{Betting earnings on \$640 (\$10 each game) using different systems and Accuracy (betting uses the best team only)}
\label{BettingEarnings}
\begin{tabular}{|l|c|c|c|}
\hline
& Betting &	Betting  & Accuracy\\ 
& (probability  & (best &\\
& distribution) & team)&\\
\hline 
\hline 
count baseline & 631.47 & 546.7 & 0.4531\\ 
\hline 
coin flip baseline & \textbf{656.60} &	513.2 & 0.4062\\ 
\hline 
ranking baseline & \textbf{642.01} &	637.7 & 0.5468\\ 
\hline 
naive ranking baseline & 637.70 &	637.7 & 0.5468\\ 
\hline 
prediction market & \textbf{640.00} & \textbf{646.9} & 0.5625\\ 
\hline 
strong predictions & 625.38 &	591.6 & 0.5156\\ 
\hline 
all predictions & 626.86 &	\textbf{651.1} & 0.5156\\ 
\hline 
support & 638.02 &	\textbf{696.9} & 0.5\\ 
\hline 
Logistic Regression & 626.54 &	610.4 & 0.5312\\ 
\hline 
SMOreg & 626.89 &	591.6 & 0.5156\\ 
\hline
\end{tabular}
\end{center}
\end{table}%

\subsection{Can we earn money in the betting market?} 
Finally, we check if our existing system, as it is, can actually earn money in the betting market. We use two approaches for betting money. Our first approach to convert our probability to betting was to bet all the money on the team with higher probability (betting all the money on the best team only). This is a strategy as crude as measuring the accuracy, since this is win-or-loose all strategy depending on the outcome 
of the game.
 We indeed find that the accuracy (percentage of the game outcomes predicted correctly) of a system has a high correlation (Pearson's coefficient $>$0.75) with betting on the best team (Table \ref{ConfusionMatrix}). 
There are subtle differences between those metrics. Betting on the best team benefits systems that predict a game with high odds (the system would earn more money), whereas accuracy gives same reward for a high odds game and a low odds game. 
The problem with both betting on the best team and accuracy is that they fail to capture the probability distribution of the predictions. For example, assume one system predicts a team will win with 51\% probability and another system predicts the same team will win with 90\% probability. If the team eventually wins the game, then both systems will get the same reward. 
Our second approach is the \emph{optimal} betting strategy according to information theory \cite{TJ_Betting}. This approach bets money on games according to a particular probability distribution, i.e. if probability of $team_1$ is x\%, then put x\% of the money for $team_1$. This way, if a system predicted 51\% probability for a team and another predicted 90\% probability for the same team, then we invest according to probability distribution and hence the earnings reflects how well the system predicted the game outcome.

We spent \$10 for each game (total of \$640). Betting all the money on the betting line baseline (after converting betting odds to probabilities) only profits \$6.9, whereas the system that performed higher according to RMSE (TwitterPaul using strong predictions) looses \$48.4. However, the system that makes use of all the predictions earns \$11.1 and the system that uses the support tweets is able to earn \$56.9. This latter method performed even worse than the naive count baseline (using RMSE) in earlier experiments. 
The system that uses the support tweets has a lower accuracy than the rest, but it is able to predict a few games with higher odds (betting more money on them). This is remarkable, since the rest of the social signal was predicting for other teams. Given that the competition only runs for 64 games, a few exceptional games are able to make a big difference in monetary performance. 

On the contrary, the optimal betting (betting using a probability distribution) has a low correlation against accuracy (Pearson's coefficient $<$0.31 against RMSE for the betting line, $<$0.16 against RMSE of the actual outcome, and $<$0.14 against accuracy). Another interesting finding in the data was that some baselines performed very well (coin flip baseline, ranking baseline). Finally, if we considered exact probabilities then prediction market baseline would eventually get exact \$640, no gain or no loss.

TwitterPaul's performance with RMSE is very close to that of the betting line (highest baseline) and better than other baselines. However, when considering the amount of money earned in the prediction market, the system is less performing. When learning the weights for individual users, we consider their performance on previous history, combining this with earnings in the betting market (capturing that high odds games result in high rewards) might help for machine learning methods to regress more accurate weights.

\begin{table*}[htdp]
\begin{center}
\caption{Confusion Matrix of Correlations between RMSE, Accuracy and Betting earnings}
\label{ConfusionMatrix}
\begin{tabular}{|l|c|c|c|c|c|c|}
\hline
& 1 - RMSE & 1-RMSE   & Accuracy& Betting  & Betting   \\
&  actual outcome & betting line & & (best team) &  (probability distribution) \\
\hline 
1 - RMSE actual & 1 &    & &  &    \\
\hline 
1-RMSE betting line  & 0.9214 &  1  & &  &    \\
\hline 
Accuracy  & -0.0212 & 0.2142   & 1 &  &    \\
\hline 
Betting (best team)  & -0.0896 & -0.0246   & 0.7543 & 1 &    \\
\hline 
Betting (probability dist.)  & 0.1575 & 0.3094  & 0.1339 & 0.1165 & 1   \\ 
\hline
\end{tabular}
\end{center}
\end{table*}%

\section{Conclusions}
\label{sec:conclusions}
In this paper we tackled the problem of predicting the outcomes of football games of the World Cup 2010 using social media chatter. We firstly developed a high-precision extraction module that scans Twitter data and parses tweets to i) detect whether they contain predictions or not and ii) classify the prediction issued in the tweet based on their modal words. The predictions extracted are mapped on to one out of the 64 games played in the World Cup, and all the predictions for a single game are aggregated in order to come up with a probability distribution of the game outcome.

We experimented with several ways of weighting the historical features for individual users, and the effect of different types of predictions. As a main conclusion, TwitterPaul is able to replicate the betting market's performance automatically, out of Twitter data. As a side result, we observed empirically that a few high-quality predictions are better than a larger number of low-confidence predictions in order to minimize root mean square error over the real outcome of the game. Furthermore, historical features, weighted in a different number of ways, did not contribute much to improve  the performance of the methods. Finally, we investigated whether the systems could earn money in the betting market, and explained why RMSE and accuracy have a low correlation with earnings. 

There are some future lines of research addressing both the extraction and prediction problem. For the former, the approach presented here is language dependent even though it is possible that the CFG extracts some predictions in languages other than English, if the team name extraction module knows how to extract the team names. There should be future work on finding language independent extraction module. We also ignore negation in this current work. For the prediction problem, an interesting line would be to detect demographic bias, to learn how to predict difficult games, and to incorporate the estimate of the odds of a game into the model. Finally, if a prediction is about winning World Cup (e.g. \textit{i predict brazil will win the world cup}), we convert it to win for next game only, but it would be interesting to see people's predictions on who will win the World Cup and how this perception changes over time. 



%
\bibliographystyle{abbrv}
\bibliography{TwitterPaul}

\end{document}